\documentclass[twocolumn,trackchanges]{aastex7}

\received{\today}
\revised{\today}
\accepted{\today}
\usepackage{amsmath}	
\usepackage{amssymb}
\usepackage[abs]{overpic}
\usepackage{xcolor,varwidth}
\usepackage[normalem]{ulem}

\newcommand{\rstar}{r_{\star}}
\newcommand{\hammer}{\texttt{H-AMR}}

\shorttitle{MHD Simulations of Giant Magnetar Flares}
\shortauthors{Chatterjee et al.}

\begin{document}

\title{Relativistic Magnetohydrodynamic Simulations of Giant Magnetar Bursts}

\author[orcid=0000-0002-2825-3590]{Koushik Chatterjee}
\affiliation{Department of Physics, University of Maryland, 7901 Regents Drive, College Park, MD 20742, USA}
\affiliation{Institute for Research in Electronics and Applied Physics, University of Maryland, College Park, MD 20742, USA}
\affiliation{Canadian Institute for Theoretical Astrophysics, 60 St. George Street, Toronto, ON M5S 3H8, Canada}
\email[show]{kchatt@cita.utoronto.ca}   

\author[orcid=0000-0001-7801-0362]{Alexander Philippov}
\affiliation{Department of Physics, University of Maryland, 7901 Regents Drive, College Park, MD 20742, USA}
\affiliation{Institute for Research in Electronics and Applied Physics, University of Maryland, College Park, MD 20742, USA}
\email{sashaph@umd.edu}   

\author[orcid=0000-0001-5660-3175]{Andrei M. Beloborodov}
\affiliation{Physics Department and Columbia Astrophysics Laboratory, Columbia University, 538 West 120th Street, New York, NY 10027}
\affiliation{Max Planck Institute for Astrophysics, Karl-Schwarzschild-Str. 1, D-85741, Garching, Germany}
\email{amb2046@columbia.edu}  

\author[orcid=0000-0001-6173-0099]{Kyle Parfrey}
\affiliation{Princeton Plasma Physics Laboratory, Princeton, NJ 08540, USA}
\email{kparfrey@pppl.gov}

\author[orcid=0000-0002-7301-3908]{Bart Ripperda}
\affiliation{Canadian Institute for Theoretical Astrophysics, 60 St. George Street, Toronto, ON M5S 3H8, Canada}
\affiliation{Department of Physics, University of Toronto, 60 St. George Street, Toronto, ON M5S 1A7, Canada}
\affiliation{David A. Dunlap Department of Astronomy, University of Toronto, 50 St. George Street, Toronto, ON M5S 3H4, Canada}
\affiliation{Perimeter Institute for Theoretical Physics, 31 Caroline St. North, Waterloo, ON N2L 2Y5, Canada}
\email[]{ripperda@cita.utoronto.ca}

\author[orcid=0000-0002-0491-1210]{Elias R. Most}
\affiliation{TAPIR, Mailcode 350-17, California Institute of Technology, Pasadena, CA 91125, USA}
\affiliation{Walter Burke Institute for Theoretical Physics, California Institute of Technology, Pasadena, CA 91125, USA}
\email[]{emost@caltech.edu}

\begin{abstract}

Gradual crustal deformation can generate strongly twisted magnetic fields around magnetars, potentially triggering giant flares with total energies exceeding $10^{44}\,\mathrm{erg}$. In this Letter, we present the first relativistic magnetohydrodynamic simulation of a surface shear-driven magnetar eruption, capturing reconnection-driven plasma heating, the ejection of relativistically hot plasma, and the formation of a hot fireball confined within the inner magnetosphere. We find that magnetic reconnection in the equatorial current sheet launches a hot trailing outflow capable of powering the initial spike observed in giant flares, while simultaneously leaving behind a thermally stratified fireball with sufficient thermal energy to produce the pulsating, decaying tail. Together, these features provide a self-consistent physical framework for understanding the observed energetics of magnetar giant flares. The eruption also expels a magnetically dominated giant plasmoid carrying up to $\sim 9\%$ of the magnetosphere’s total magnetic energy. Furthermore, our simulation demonstrates how the plasmoid drives the formation of a blast wave---an important ingredient in models linking magnetar eruptions to fast radio bursts.

\end{abstract}


\section{Introduction} 

Magnetars are young neutron stars with magnetic fields exceeding $10^{14}\,$G and rotation periods of several seconds \citep{Duncan:1992}. They reveal themselves through sporadic high-energy activity, the most dramatic of which are \emph{giant flares} \citep[for a review, see][]{Kaspi:2017}. To date, three confirmed giant flares have been detected: in 1979 from the soft gamma repeater (SGR) 0526--66 \citep{Evans:1980}, in 1998 from SGR 1900+14 \citep{Hurley:1999}, and in 2004 from SGR 1806--20 \citep{Hurley:2005}. Each event began with an intense, sub-second spike of hard X-rays and gamma rays, followed by a tail of softer X-rays modulated by the magnetar’s rotation period. Magnetars have also been closely linked to fast radio bursts (FRBs), with the Galactic magnetar SGR 1935+2154 exhibiting coincident X-ray flaring and FRB emission \citep[][]{Bochenek:2020, Chime:2020, Mereghetti:2020}. 

The energy source for magnetar bursting activity is believed to be the evolution of the star’s internal magnetic field, which induces stresses in the crust, leading to surface motions and the twisting of magnetospheric field lines. The twisting motion may occur rapidly, generating Alfv\'en and fast magnetosonic waves that propagate into the magnetosphere \citep{Yuan:2020, Burnaz:2025}, or proceed slowly, storing magnetic energy until it is released in a sudden, violent eruption \citep{Parfrey:2012}. 

Here we focus on the latter scenario. Models of slowly twisted magnetospheres in the framework of force-free electrodynamics (FFE) have shown that both axisymmetric and fully three-dimensional twists can drive eruptive behavior when the total twist angle exceeds a critical threshold $\psi_{\rm crit} \gtrsim \pi$ \citep[][]{Uzdensky:2002,Parfrey:2013,Carrasco:2019, Mahlmann:2023}. Although FFE simulations successfully show that a significant fraction of the magnetic twist energy, $\mathcal{E}_{\rm twist}$, can be released during such eruptions, the method 
is unable to track the dissipated energy deposited as heat into the plasma, which is crucial for understanding the observed emission. The problem requires the full framework of relativistic magnetohydrodynamics (MHD) that self-consistently evolves both the electromagnetic fields and the plasma.

In this \emph{Letter}, we present the first relativistic MHD simulation of a magnetar giant flare, focusing on the structure, evolution and energetics of the flaring magnetosphere. Our axisymmetric simulation shows how magnetic reconnection launches ultrarelativistic magnetized ejecta and loads the inner closed magnetosphere with hot plasma, creating a trapped ``fireball''. We measure the basic parameters of the flare, including the partitioning of the released energy between the trapped fireball and the ejecta, thermal and electromagnetic, which are expected to power the gamma-ray flare and a possible FRB.

\begin{figure*}
    \centering
    \includegraphics[width=\textwidth,trim= 60mm 0mm 50mm 0mm, clip]{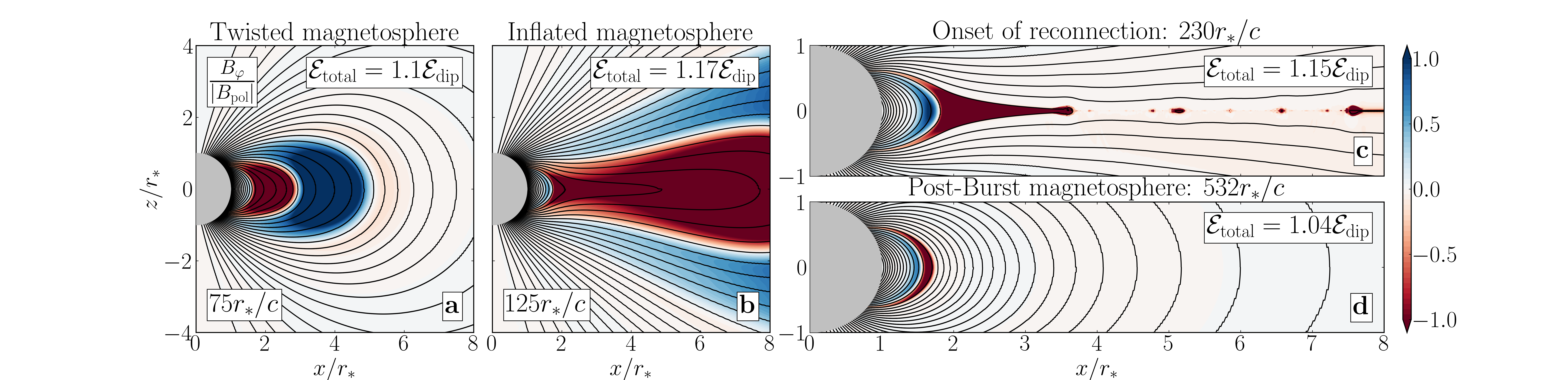}
    \includegraphics[width=\textwidth,trim= 65mm 0mm 70mm 0mm, clip]{FIG2.png}
    \caption{Onset and evolution of the eruption. \textit{Top row (a–d):} Evolution of the inner magnetosphere ($r \leq 30\,\rstar$) during a flare, showing the toroidal-to-poloidal magnetic energy ratio and the total magnetic energy. The magnetosphere initially inflates as the rotating stellar surface twists the magnetic field lines (a–b). Following magnetic reconnection in the equatorial plane (c), the magnetosphere relaxes toward a nearly dipolar configuration, retaining only a small residual toroidal field (d). \textit{Bottom row (e–g):} Time evolution of the plasma internal energy, $U$, in the inner magnetosphere, illustrating the formation of a hot, magnetized fireball. The internal energy is normalized to the magnetic energy density $U_{\rm B_*}=B_*^2/8\pi$ corresponding to the stellar magnetic field $B_*$ (Eq.~\ref{eqn:edip}); the magenta line marks the reconnecting field line. The hot magnetization $\sigma_{\rm hot}$ (panel \textit{h}) remains above unity within the fireball, indicating that the magnetosphere remains magnetically dominated.
    }
    \label{fig:inner}
\end{figure*}

\section{\label{sec:method}Numerical Methods}

The initial unperturbed magnetosphere at time $t=0$ is set up as
a static dipolar magnetic field anchored to the star of radius $r_\star$. The magnetic field strength is determined by a chosen field $B_*$ at the magnetic poles of the star. The total energy of the dipole magnetosphere is
\begin{equation}\label{eqn:edip}
    \mathcal{E}_{\rm dip} \approx 3\times 10^{47} \left( \frac{B_*}{10^{15}\,{\rm G}} \right)^2 \left( \frac{\rstar}{10\,{\rm km}} \right)^3\ {\rm erg}.
\end{equation}
The initial magnetosphere is filled with a static plasma of rest-mass density $\rho$ and a small pressure $p$. It has a uniform cold-plasma magnetization parameter
$\sigma_{\rm cold} \equiv b^2 / 4\pi \rho c^2 = 500$ 
and plasma-beta parameter $\beta \equiv 8\pi p/b^2= 10^{-8}$, where $b^2/8\pi$ is the magnetic energy density measured in the rest-frame of the MHD fluid. Here,
$b^2 = b^{\mu} b_{\mu}$ where $b^{\mu}$ is the contravariant four-vector constructed using the fluid four-velocity $u^{\mu}$, the electromagnetic field tensor $F^{\mu\nu}$, and the Levi--Civita tensor:
\begin{equation}
    b^{\mu}\equiv \frac{1}{2}\epsilon^{\mu\nu\kappa\lambda}u_{\nu}F_{\lambda\kappa}.
\end{equation}

At time $t>0$, the magnetosphere is gradually twisted by driving an axisymmetric shear motion of its footpoints on the star. We prescribe a shear profile similar to \cite{Parfrey:2013} and  \cite{Mahlmann:2023}:
\begin{equation}\label{eqn:twist}
    \omega(\theta)=\frac{\omega_0\sin[(\theta-\theta_{\rm C})(\pi/\theta_{\rm T})]}{1+\exp{[\kappa(|\theta-\theta_{\rm C}|-\theta_{\rm T})]}},
\end{equation}
where the sine factor provides surface velocities that change direction across $\theta = \theta_{\rm C}$, thereby producing a shear twist. Our fiducial simulation places the twist center at co-latitude $\theta_{\rm C} = 45^{\circ}$ and adopts a fall-off factor $\kappa = 50$ in the sheared stripe of angular half-width $\theta_{\rm T} = 0.05\pi$.\footnote[1]{We also explore alternative twist locations to investigate their impact on the eruption dynamics, see Appendix~\ref{sec:twist_loc}.}
The twisting rate is controlled by the parameter $\omega_0$. It is linearly ramped up from zero
to $\omega_0 = 0.04\, c/\rstar$ at the beginning of the simulation and then held constant until the magnetic reconnection flare is triggered at time $t_{\rm rec}$; we ramp-down $\omega_0$ at $t_{\rm rec}$. The chosen value of $\omega_0$ is sufficiently small to weakly affect the total twist energy at the eruption onset, while remaining large enough to mitigate losses of the magnetospheric twist due to numerical diffusion \citep{Mahlmann:2023}. 

The plasma is strongly heated during the magnetospheric flare, and the plasma equation of state becomes important in the heated regions. We adopt an equation of state that relates the fluid pressure $p$ to the internal energy density $U$ as $p = (\Gamma_{\rm ad} - 1)U$, with an adiabatic index $\Gamma_{\rm ad} = 4/3$. This choice is appropriate for the radiation-dominated electron--positron plasma produced in the reconnection region. The plasma is assumed to be optically thick, so that radiation diffusion can be neglected over the duration of the simulation, $t \sim 25\,\mathrm{ms}$. As a result, the generated heat is trapped and advected with the plasma.

The evolving and erupting magnetosphere also has cold regions, which are not heated by magnetic reconnection. In these regions, the FFE approximation would be adequate and useful, as it helps avoid spurious heating that appears in MHD simulations at very low plasma-$\beta$ and high magnetization $\sigma_{\rm cold}$. We maintain stable FFE-like conditions in these regions by enforcing a pressure floor and a density floor, which correspond to a minimum $\beta=10^{-8}$ and a maximum $\sigma_{\rm cold}=500$. 

The floors ensure that the cold plasma remains magnetically dominated, while plasma heated by magnetic reconnection is evolved using the full MHD equations without any floor treatment (see Appendix~\ref{sec:FFE} for details).\footnote[2]{Our hybrid FFE--MHD implementation follows two complementary approaches: (i) during the pre-burst phase, we evolve the energy equation and impose floor values on the magnetization and plasma-$\beta$ \citep{Parfrey:2017,Parfrey:2024}; and (ii) during the flare and post-burst phases, we select the primitive-variable recovery method---either entropy inversion \citep{Noble:2009, Liska:2022:hamr} or energy inversion---based on local plasma conditions.}
The numerical stability and accuracy of the simulation is also improved by suppressing the velocity parallel to the magnetic field for the floored cold plasma in the vicinity of the star \citep[e.g.,][]{Tchekhovskoy:2013}. Our method is able to conserve energy in the simulation domain with error $\lesssim0.1\%\,\mathcal{E}_{\rm dip}$.

We use the GPU-accelerated code \hammer{} \citep{Liska:2022:hamr} to solve the general relativistic (GR) MHD equations in logarithmic spherical polar coordinates $(t, \log r, \theta, \varphi)$. We adopt the Minkowski metric, neglecting gravity. Grid length scales are measured in units of the magnetar radius, $\rstar$, and timescales in $\rstar/c$. Our 2D simulations use an effective resolution of $20160 \times 20160$ cells, including a single layer of mesh refinement within $15^\circ$ of the equatorial plane. The computational domain extends from $r \in [1, 750]\,\rstar$ and $\theta \in [0, \pi]$. We adopt reflective boundary conditions (BCs) at the poles and at the stellar surface, periodic BCs in the $\varphi$-direction, and outflow BCs at the outer radial boundary. We fix the radial magnetic field $B^r$ at the surface to be constant, and extrapolate the magnetic field components into the star (i.e., the ghost zone cells) as $(B^r,\,B^{\theta},\,B^{\varphi})\propto (r^{-3}, r^{-4}, r^{-2})$, where $B^i$ are the contravariant field components in spherical coordinates. From hereon, we express vector components in the orthonormal basis unless otherwise stated.

\begin{figure*}
    \centering
    \includegraphics[width=0.49\textwidth,trim= 8mm 0mm 25mm 20mm, clip]{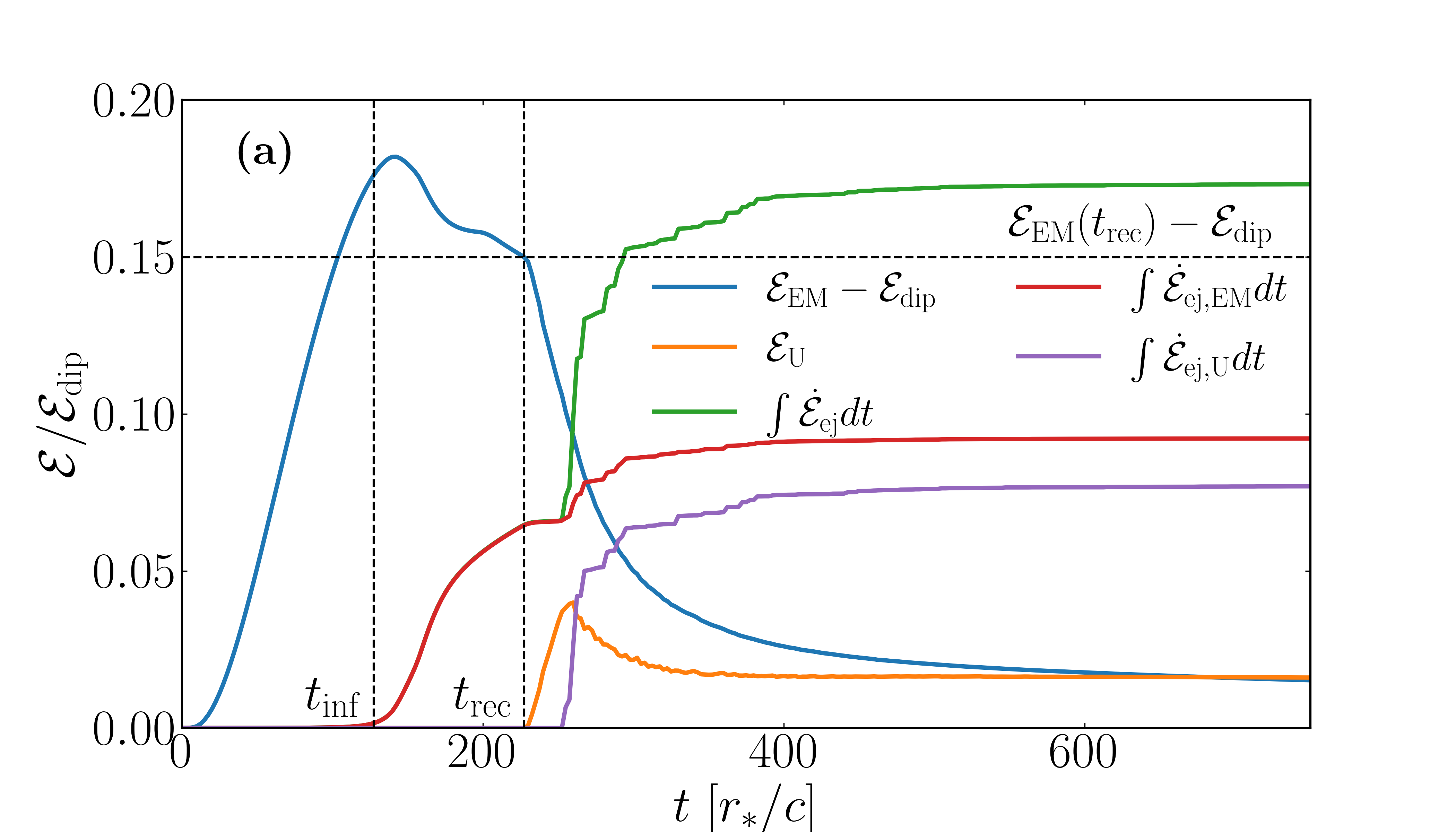}
    \includegraphics[width=0.49\textwidth,trim= 15mm 0mm 25mm 20mm, clip]{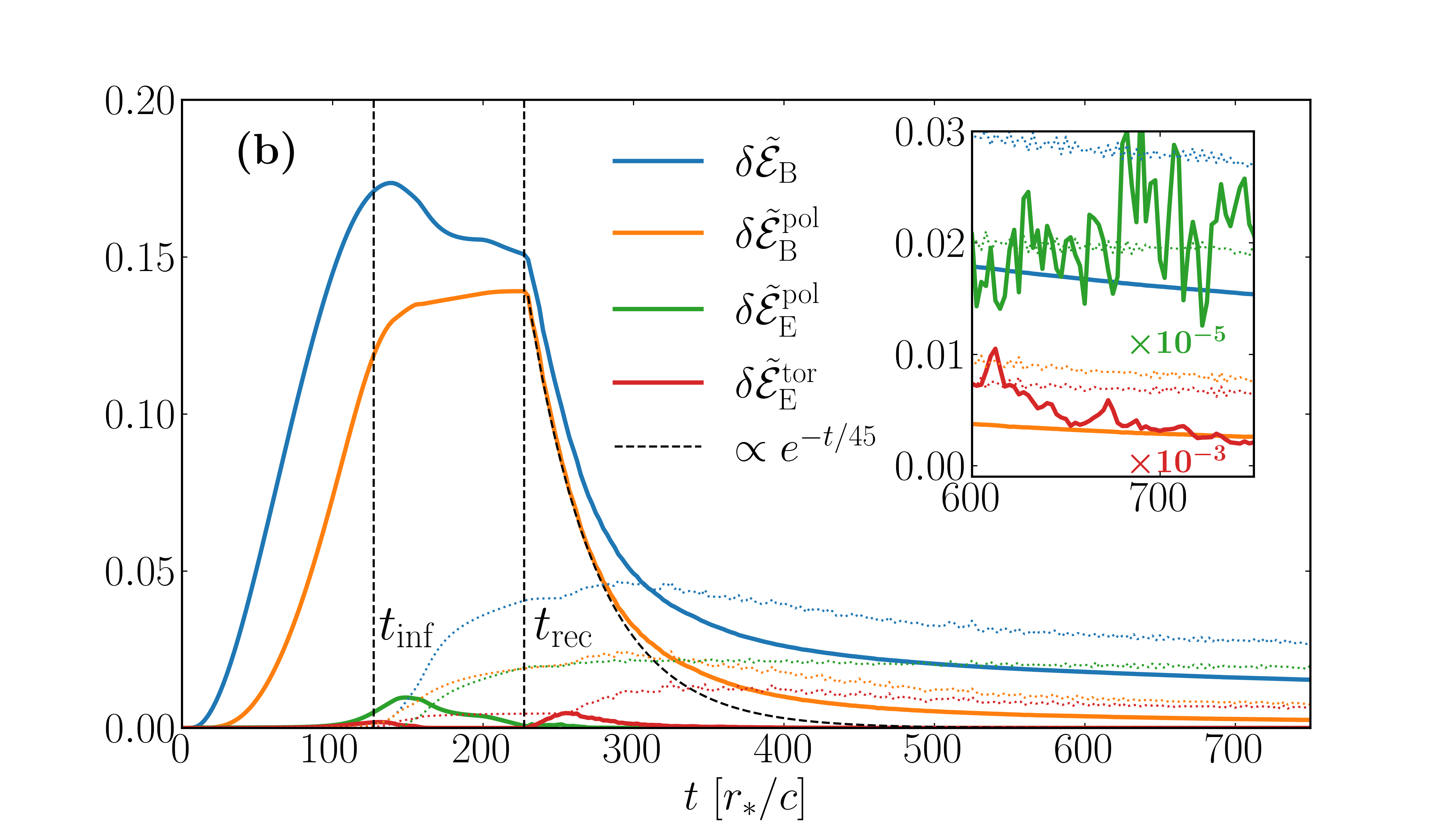}
    
    \caption{Flare energetics. \textit{(a)} Inner magnetosphere: electromagnetic energy, $\mathcal{E}_{\rm EM}$; thermal energy, $\mathcal{E}_{\rm U}$; and the time-integrated ejected energy, $\int \dot{\mathcal{E}}_{\rm ej}dt$, together with its individual components. We find that nearly $17\%$ of the dipolar magnetic energy, $\mathcal{E}_{\rm dip}$, is released as ejecta, powering the flare, while a substantial amount of hot plasma remains trapped in the inner magnetosphere. The horizontal dashed line denotes the twist energy stored in the inner magnetosphere at the onset of reconnection, and the vertical dashed lines mark the times of inflation and reconnection onset, respectively. \textit{(b)} Decomposition of the electromagnetic energy in the inner magnetosphere (solid curves) and outside the inner region ($r > 30r_*$; dotted curves). Shown are the energies of the total and poloidal components of the magnetic field, $\mathcal{E}_{\rm B}$ and $\mathcal{E}_{\rm B}^{\rm pol}$, as well as the poloidal and toroidal electric field components, $\mathcal{E}_{\rm E}^{\rm pol}$ and $\mathcal{E}_{\rm E}^{\rm tor}$. In each case, we plot the fractional change $\delta \tilde{\mathcal{E}} \equiv (\mathcal{E} - \mathcal{E}_{t=0}) / \mathcal{E}_{\rm dip}$. The dashed curved line indicates an exponential fit to the decay of $\mathcal{E}_{\rm B}^{\rm pol}$ due to reconnection. The inset shows a zoomed-in view at late times.}
    \label{fig:energetics}
\end{figure*}

\section{\label{sec:results}Results}

The surface shear twists the dipolar magnetic field lines, generating a toroidal component (Fig.~\ref{fig:inner}a) that grows as $B_{\varphi} \approx (\omega_0 t)\, B_{\rm pol}$. 
At $t_{\rm inf} \approx 120\,\rstar / c$, the initially dipolar poloidal structure of the magnetosphere begins to inflate toward a split-monopole configuration (Fig.~\ref{fig:inner}b). The inflated magnetic field lines have footprints on the star at $0<\theta\lesssim \theta_{\rm C}$; they form the outer magnetosphere that is pushed outward by the increasing pressure $B_{\varphi}^2/8\pi$ of the twisted field lines with footprints at $\theta\sim\theta_{\rm C}$, where surface shear is applied. We find that significant inflation occurs when the total twist angle reaches $\psi_{\rm crit} \approx \omega_0 t_{\rm inf} \gtrsim \pi$, consistent with previous measurements of the critical angle $\psi_{\rm crit}$ \citep[e.g.,][]{Mahlmann:2023}. 

At $t = t_{\rm rec} \approx 215\,\rstar / c$, the inflated field lines become sufficiently stretched to form an equatorial current sheet and trigger magnetic reconnection (Fig.~\ref{fig:inner}c). To isolate the energetics of a single ejection event, we terminate the surface twisting at the onset of reconnection in the inflated magnetosphere (continued twisting would result in repeated eruptions, see \citealt{Parfrey:2013}). At $t>t_{\rm rec}$, the outer parts of the strongly inflated field lines pinch off via magnetic reconnection while their inner parts snap back toward the star, restoring the dipolar configuration with a small residual twist (Fig.~\ref{fig:inner}d). The reconnection occurs at multiple X-points in the equatorial current sheet, producing multiple hot plasmoids (Fig.~\ref{fig:inner}e) which are expelled at relativistic speeds from the reconnection site. This process detaches the inflated outer magnetic loops from the star. The resulting ejecta contains a giant magnetized plasmoid composed of closed poloidal field lines filled with toroidal magnetic field (see Fig.~\ref{fig:bubbletail}).

Note that only field lines with footprints around $\theta_{\rm C}=45^\circ$ are strongly twisted, and field lines anchored at smaller $\theta$ are twist-free. The reconnection process first develops in the twisted field-line bundle and later involves the twist-free field lines. During the later phase, $B_{\varphi}\approx0$ on the reconnecting field lines, which means there is no guide field in the reconnection layer.

Some plasmoids created in the reconnection layer move inward and join the closed magnetosphere. This hot plasma descends toward the stellar surface, reflects off it, and eventually settles into a quasi-stationary layer of hot gas trapped in the closed magnetic loop
(Fig.~\ref{fig:inner}f). As reconnection proceeds, newly closed magnetic field lines loaded with hot plasma form successively and build up a stratified structure in which the internal energy decreases outward, forming a hot, magnetized ``fireball'' confined by the intense magnetic field of the magnetar (Fig.~\ref{fig:inner}g). The fireball remains magnetically dominated, with a ``hot magnetization'' parameter $\sigma_{\rm hot} \equiv b^2/4\pi  (\rho c^2 + U + p) \gtrsim 10$ (Fig.~\ref{fig:inner}h). 

\begin{figure*}
    \centering
    \includegraphics[height=0.29\textwidth,trim= 0mm 0mm 10mm 10mm, clip]{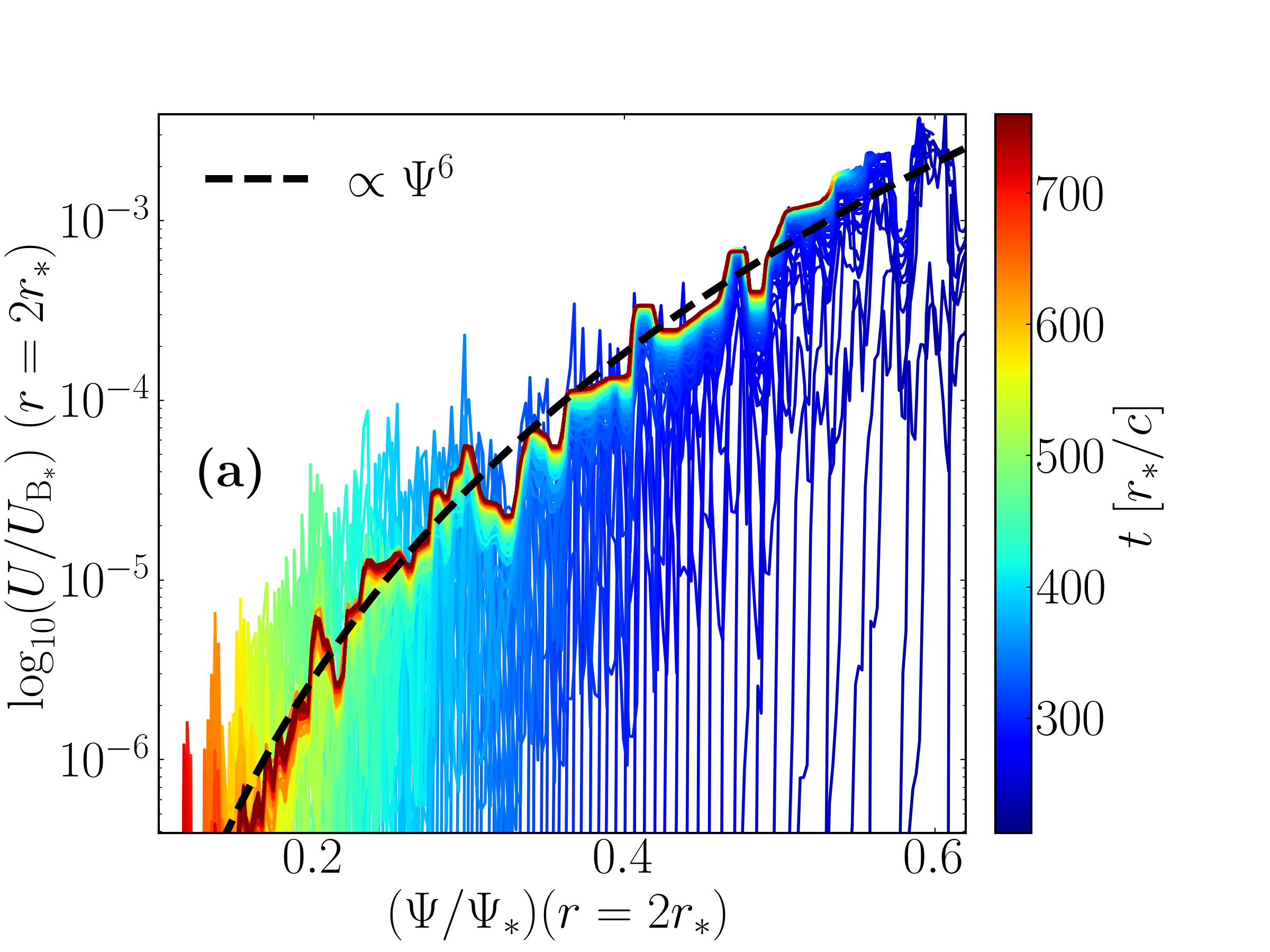}
    \includegraphics[height=0.29\textwidth,trim= 5mm 0mm 10mm 10mm, clip]{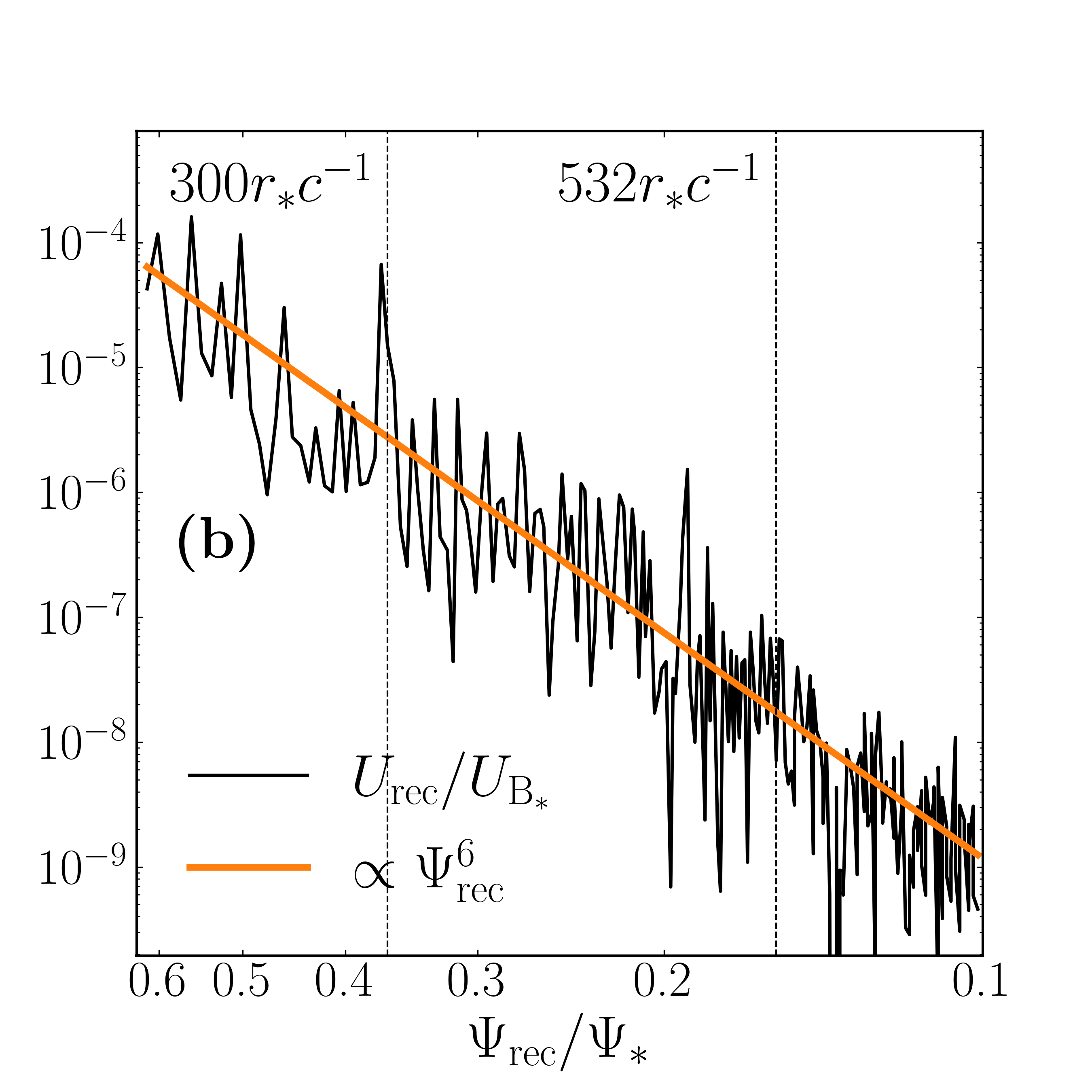}
    \includegraphics[height=0.29\textwidth,trim= 5mm 0mm 10mm 10mm, clip]{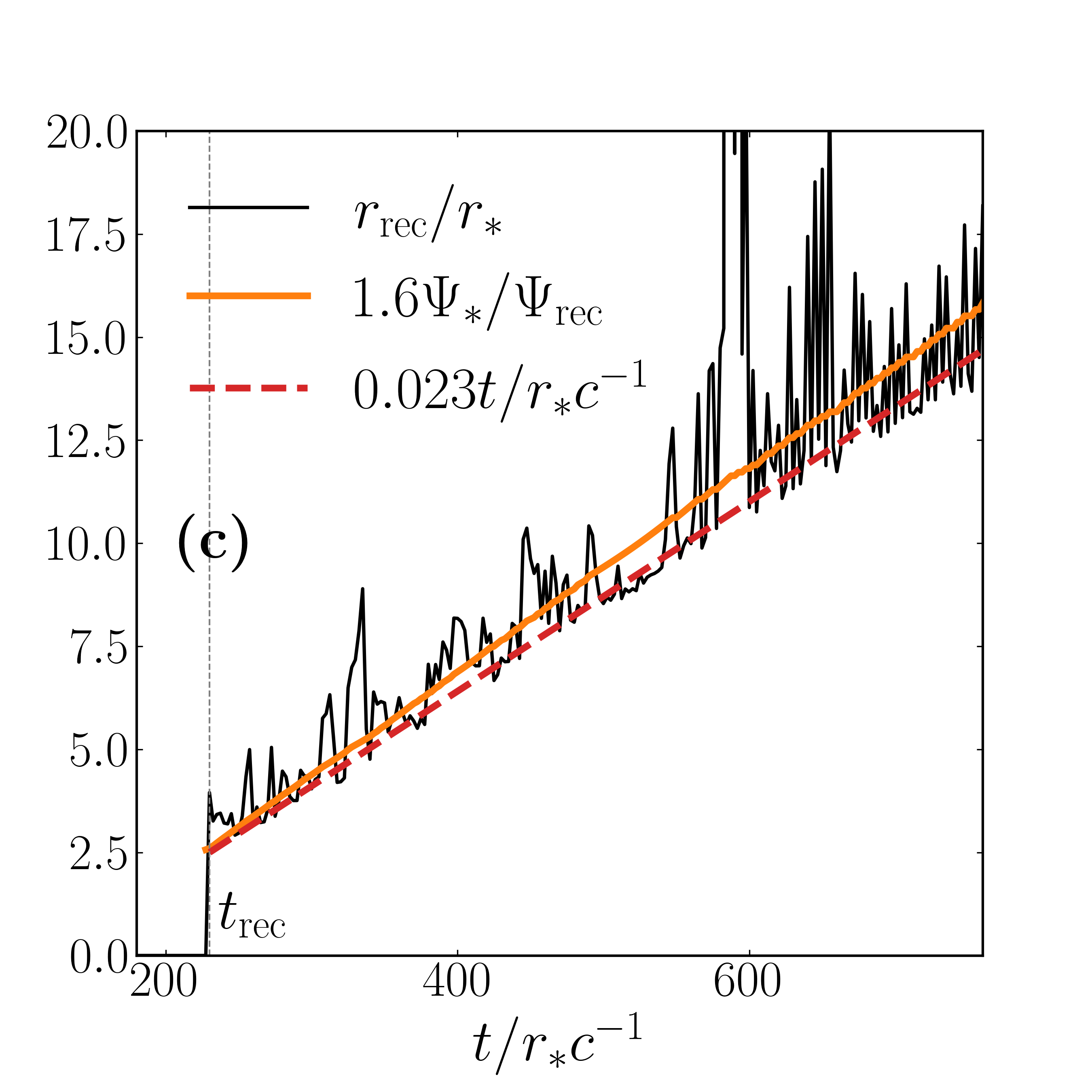}
    \caption{Time evolution of the plasma internal energy, $U$, and its connection to the history of reconnection. \textit{(a)} Time evolution of $U$ as a function of the poloidal flux function, $\Psi$, both evaluated at a fixed radius of $2\,\rstar$. At late times, $U \propto \Psi^6$, as expected for a nearly dipolar magnetic field at the reconnection radius (see Sec.~\ref{sec:fireball}). $U$ is normalized by $U_{\rm B_*}$ (see Fig.~\ref{fig:inner}), and $\Psi$ is normalized by the total stellar magnetic flux, $\Psi_*$. \textit{(b)} Internal energy, $U$, as a function of the flux function evaluated at the reconnection radius, $\Psi_{\rm rec}$, demonstrating $U \propto \Psi^6$, consistent with the nearly dipolar magnetic field at the reconnection radius and with panel (a).  \textit{(c)} Temporal evolution of the reconnection radius, $r_{\rm rec}$ (black curve), which migrates outward over time. Its evolution closely tracks that of $1/\Psi_{\rm rec}$ (orange curve). A simple fit-by-eye to $r_{\rm rec}(t)$ is overplotted, providing an estimate of the reconnection rate (see Sec.~\ref{sec:fireball}).
    }
    \label{fig:rec}
\end{figure*}

\subsection{Flare Energetics}

The total energy budget of the flare is set by the energy $\mathcal{E}_{\rm twist}$ injected into the magnetosphere during the gradual twisting process. In general, $\mathcal{E}_{\rm twist}$ required for eruption depends on the location and profile of the surface shear \citep{Parfrey:2013, Mahlmann:2023}. In our fiducial simulation, $\mathcal{E}_{\rm twist} \approx 22\%\,\mathcal{E}_{\rm dip}$ triggers the flare. In this subsection, we analyze how the energy stored in the inflated magnetosphere is
partitioned between the energy carried by the ejecta and the residual thermal energy
trapped in the closed magnetosphere.

During the time interval $t_{\rm inf} \lesssim t \lesssim t_{\rm rec}$, the poloidal field rapidly inflates toward a split--monopole configuration, $B_{\rm pol} \propto r^{-2}$. We choose a sphere of radius $r=30r_\star$ to measure the outflow of electromagnetic energy during the inflation and later phases.
After the onset of reconnection, the magnetically dominated ejecta fully decouple from the magnetosphere and escape to infinity. Within the reconnecting current sheet, magnetic energy stored in the inflated magnetosphere is converted into thermal and kinetic energy of the plasma.

To quantify the partitioning of energy in the erupting magnetosphere, we analyze the evolution of the electromagnetic energy $\mathcal{E}_{\rm EM}$ and the thermal energy $\mathcal{E}_{\rm U}$ contained in the region
$r < 30\,r_*$:
\begin{equation}
    \mathcal{E}_{\rm EM, U}=\int_{r<30\rstar} 
   T^{tt}_{\rm EM, U}
    \,\,dV, 
\end{equation}
where $T_{\rm EM}$ and $T_{\rm U}$ are the electromagnetic and thermal components of the total stress--energy tensor of the MHD,
$T^{\mu}_{\,\,\,\nu} \equiv (\rho c^2 + U + p + b^2/4\pi)\,u^\mu u_\nu
+ (p + {b^2}/{8\pi})\delta^\mu_{\,\,\nu}
- {b^\mu b_\nu}/{4\pi},$
and $dV=r^2\sin{\theta}\,dr\,d\theta \, d\varphi$ is the volume-element in spherical coordinates $(t,r,\theta,\varphi)$. In addition, we decompose ${\cal E}_{\rm EM}$ into two parts, $\mathcal{E}^{\rm pol}$ and $\mathcal{E}^{\rm tor}$, carried by the poloidal and toroidal field components. We also track the energy ejected through the sphere of radius $r = 30 \rstar$ by calculating the integral
\begin{equation}
\mathcal{E}_{\rm ej} \equiv \int \dot{\mathcal{E}}_{\rm ej} dt,   \quad
\dot{\mathcal{E}}_{\rm ej} = \iint T^r_{\,\,t}\,\, r^2\sin{\theta}\, d\theta d\varphi \big|_{r = 30 \rstar}.
\end{equation}

Figure~\ref{fig:energetics} shows the evolution of the different components of the total energy remaining inside $30r_\star$, as well as the energy components ejected through the sphere of $r=30r_\star$. We find that the ejecta carries away most of the twist
energy, $\mathcal{E}_{\rm ej} \approx 17.5\%\,\mathcal{E}_{\rm dip}
\simeq 80\%\,\mathcal{E}_{\rm twist}$ (Fig.~\ref{fig:energetics}a, green line). The ejecta energy is almost equally divided between electromagnetic field,
$\mathcal{E}_{\rm ej,\,EM} \approx 9\%\,\mathcal{E}_{\rm dip}$ (primarily stored in the escaping giant plasmoid), and heat, $\mathcal{E}_{\rm ej,\,U} \approx 7.7\%\,\mathcal{E}_{\rm dip}$ (contained in reconnection-heated plasma in the tail behind the giant plasmoid). The heated part of the ejecta can radiate the main spike of the gamma-ray flare.

Most of the magnetic twist is expelled with the ejecta. The magnetosphere inside $30r_\star$ relaxes to a slightly twisted dipolar magnetosphere, with energy exceeding $\mathcal{E}_{\rm dip}$ by only $\sim 1\%$ (Fig.~\ref{fig:inner}d, Fig.~\ref{fig:energetics}b).\footnote[3]{Once the magnetosphere has been cleared of the hot plasma and radiation generated by the flare,
resistive dissipation of the residual toroidal magnetic field can sustain the quiescent
X-ray emission of the magnetar \citep{Beloborodov:2007}.} A large part of the closed magnetosphere ends up filled with hot plasma. The thermal energy in this trapped
fireball (Fig.~\ref{fig:energetics}a, orange line) asymptotes
to $\mathcal{E}_{\rm U} \approx 1.5\%\,\mathcal{E}_{\rm dip}\approx 6.8\%\,\mathcal{E}_{\rm twist}$, which is sufficient to power the post-peak emission tail modulated with the magnetar's rotation period
\citep{Thompson:1995}. FFE simulations previously estimated that the energy dissipated during an eruption is $\mathcal{E}_{\rm U, \, total}\approx 10\%\, \mathcal{E}_{\rm twist}$ \citep{Parfrey:2013}, with an estimated upper bound of $\mathcal{E}_{\rm U, \, total}\lesssim 25\%\, \mathcal{E}_{\rm twist}$ \citep{Mahlmann:2023}. We find that the total thermal energy production in our MHD simulation is $\mathcal{E}_{\rm U, \, total}=\mathcal{E}_{\rm U, \, fireball}+\mathcal{E}_{\rm ej,\,U}=(7.7\%+1.5\%)\,\mathcal{E}_{\rm dip}\approx 42\%\,\mathcal{E}_{\rm twist}$, considerably larger than the FFE dissipation estimates.

\subsection{\label{sec:fireball}Formation of the Stratified Fireball}

It is convenient to label magnetic field lines using the poloidal magnetic flux function,
$\Psi(r,\theta) = \int_{0}^{2\pi}\!\!\int_{0}^{\theta} B^{r}\,r^2\sin{\theta}\, d\theta\, d\varphi$, since the field lines of an axisymmetric magnetosphere lie on surfaces of constant $\Psi$. In the evolving magnetosphere (with magnetic field footpoints frozen into the static star), each magnetic flux surface retains the value of $\Psi$ that it had in the initial dipole configuration. In particular, $\Psi = 0$ on the magnetic axis (which extends to infinity), and $\Psi$ increases for field lines that close progressively closer to the star, from the pole toward the equator.

Heat in the trapped fireball is generated by magnetic reconnection, so the fireball occupies flux surfaces $\Psi_{\rm rec} < \Psi < \Psi_1$ that have passed through the reconnection layer. Thermal energy is first deposited on the flux surfaces that reconnect earliest, at $\Psi_1 \approx 0.62\,\Psi_{*}$, where $\Psi_*=(1/2)\int_{r=\rstar} |B^r|r^2\sin{\theta}\,d\theta\, d\varphi$ is the total stellar magnetic flux. Heat
is added layer by layer as reconnection proceeds to $\Psi_{\rm rec}(t)<\Psi_1$. This process produces a stratified fireball characterized by a thermal pressure $p(\Psi)$ and a thermal energy density $U = 3p$, both of which are approximately uniform on each closed flux surface $\Psi$, as the fireball relaxes to pressure equilibrium (Fig.~\ref{fig:rec}a).

One can see the reconnecting flux surfaces in Fig.~\ref{fig:inner}(e--g) in representative time snapshots. Reconnection involves tearing (X-points) of the current sheet and creates closed contours of $\Psi$, i.e. flux surfaces of toroidal shape, enclosing plasmoids (Fig.~\ref{fig:inner}e). In practice, we identify plasmoids by searching for additional closed contours of $\Psi$ along the equatorial current sheet, excluding those associated with the closed magnetosphere and the ejecta. We then define the reconnection radius, $r_{\rm rec}$, as the radial position of the region between the closed zone and the nearest plasmoid identified from the $\Psi$ contours, i.e., the X-point closest to the star (see, e.g., the midplane region at $r\sim4\,\rstar$ in Fig.~\ref{fig:inner}e). At each time $t$, we define $\Psi_{\rm rec}(t)$ as the magnetic flux surface passing through this X-point. Radius $r_{\rm rec}(t)$ represents the evolving radial extension of the reconnected magnetosphere, and $\Psi=\Psi_{\rm rec}(t)$ defines the evolving outer boundary of the trapped fireball. We also measure the thermal energy density $U_{\rm rec}(t)$ at point $r_{\rm rec}(t)$; it represents the energy density being deposited into the fireball.

The simulation shows how new layers are added to the trapped fireball and how $r_{\rm rec}(t)$ moves outward while $\Psi_{\rm rec}(t)$ and $U_{\rm rec}(t)$ decrease with time (Fig.~\ref{fig:rec}b,c). The
magnetosphere at $r < r_{\rm rec}$ is approximately dipolar which gives $\Psi_{\rm rec} \propto r_{\rm rec}^{-1}$ and $B_{\rm rec}\propto r_{\rm rec}^{-3}$. As a result, the energy density generated by reconnection at the boundary radius $r_{\rm rec}$ scales as $U_{\rm rec} \propto B_{\rm rec}^2 \propto r_{\rm rec}^{-6} \propto \Psi_{\rm rec}^6$. This estimate is consistent with the evolution of $U_{\rm rec}$ observed in the simulation, as shown in Fig.~\ref{fig:rec}a, and explains the steep stratification of thermal energy density in the trapped fireball, $U(\Psi)$.

The trapped fireball grows in size at the rate at which reconnection proceeds in the current sheet. The reconnection rate $v_{\rm rec}$ in highly magnetized plasmas ranges from $\sim 0.01c$ in collisional MHD fluids to $\sim 0.1c$ in collisionless plasmas \citep[e.g.,][]{Bransgrove:2021}. Here we estimate $v_{\rm rec}$ using three complementary methods. First, we infer $v_{\rm rec}$ from the decay timescale $\tau_{\rm rec}$ of the poloidal magnetic energy shown in Fig.~\ref{fig:energetics}(b). The poloidal magnetic energy decays as $\delta \tilde{\mathcal{E}}_{\rm B}^{\rm pol} \propto e^{-t/(45\,\rstar/c)} = (e^{-t/\tau_{\rm rec}})^2$, which implies $\tau_{\rm rec} \approx 2 \times 45\,\rstar/c = 90\,\rstar/c$. The initial reconnection episode occurs at an average radius $\langle r_{\rm rec} \rangle \approx 3.5\,\rstar$, yielding a reconnection rate $v_{\rm rec} \simeq \langle r_{\rm rec} \rangle / \tau_{\rm rec} \approx 0.038c$ (Appendix\ref{sec:convergence}). Second, using the time evolution of $r_{\rm rec}$ (Fig.~\ref{fig:rec}c), we obtain a conservative lower bound $v_{\rm rec} \gtrsim \delta r_{\rm rec} / \delta t \approx 0.023c$. Finally, we directly measure the inflow speed of the upstream plasma into the current sheet by sampling the vertical velocity $v_z$ a few grid cells above the current sheet at multiple radii. After averaging over time, radius, and altitude, we find $v_{\rm rec} \approx 0.03c$. Overall, the measured reconnection rates are consistent with those reported in previous (GR)MHD simulations, which typically find $v_{\rm rec} \approx 0.01c$--$0.03c$ \citep{Ripperda:2020, Bransgrove:2021, Grehan:2025}.

\begin{figure*}
    \centering
    \includegraphics[width=\textwidth,trim= 40mm 0mm 50mm 20mm, clip]{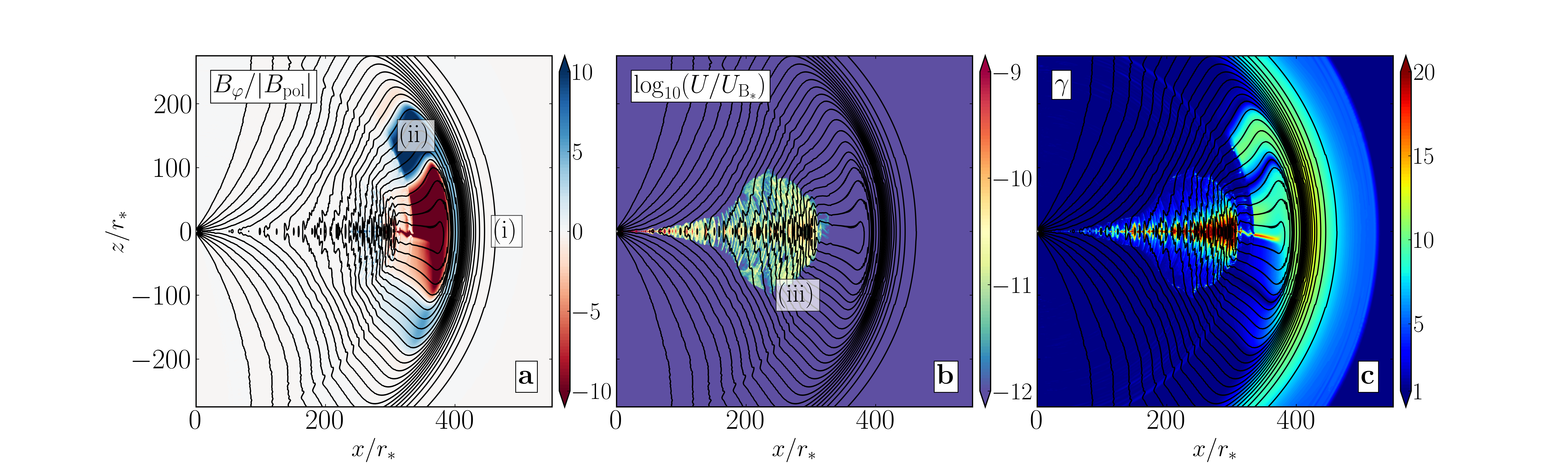}
    \caption{Structure of the flare ejecta at $532r_*/c$. Three distinct regions emerge: (i) the magnetic pulse, a relativistically moving, near-spherical envelope of compressed outer magnetospheric field lines filled with cold plasma and containing negligible toroidal field; (ii) the piston, a relativistically moving ejecta head containing cold plasma and significant toroidal magnetic fields; and (iii) the ejecta tail, filled with relativistically hot plasma moving at high Lorentz factor, which can produce the initial hard X-ray spike observed in magnetar giant flares. \textit{(a)} Ratio of toroidal to poloidal magnetic field components, highlighting the dominance of $B_{\varphi}$ in the piston. \textit{(b)} Internal energy, $U$, normalized by the stellar magnetic energy density $U_{\rm B_*}$, showing the hot plasma in the ejecta tail. \textit{(c)} Lorentz factor, $\gamma$, of the different ejecta components.}
    \label{fig:bubbletail}
\end{figure*}

\subsection{\label{sec:bubble}Magnetic Flare Ejecta}

Following the onset of reconnection, the magnetosphere ejects a giant magnetized plasmoid\footnote[4]{We find that the radial width of the 
accelerating plasmoid, $\Delta_{\rm w}$,
remains approximately constant while its transverse size, $\Delta_\perp$, grows due to sideways expansion $\Delta_\perp\propto r$. This evolution is
consistent with energy conservation, ${\cal E}_{\rm ej}\sim r^2\Delta_{\rm w}B^2$, and magnetic flux conservation, $\Phi\sim r\Delta_{\rm w}B$.}\citep[see, e.g.,][]{Beloborodov:2009, Parfrey:2013}. Field lines in the plasmoid were initially connected to the stellar twisting region prior to reconnection, so the plasmoid is $B_{\varphi}$-dominated \citep[Fig.~\ref{fig:bubbletail}a; e.g.,][]{Parfrey:2013,Barkov:2022,Mahlmann:2023}. It remains cold (Fig.~\ref{fig:bubbletail}b) and flies outward at nearly the speed of light. It acts as a piston on the outer (untwisted) magnetosphere, driving a near-spherical envelope of compressed outer magnetospheric field lines, and forms a relativistically moving magnetic pulse or blast wave \citep[Fig.~\ref{fig:bubbletail}c;][]{Beloborodov:2020, Lyubarsky:2020}. The tail of the ejecta is continuously fed by the hot plasma from the reconnecting current sheet (Fig.~\ref{fig:bubbletail}b). 

Prior to the onset of reconnection ($t \le t_{\rm rec}$), we evolve the cold plasma in the inflating outflow under enforced force-free conditions (see Appendix~\ref{sec:FFE}), maintaining a fixed cold magnetization of $\sigma_{\rm cold}=500$. During this pre-reconnection phase, the leading edge of the inflated region accelerates linearly, reaching Lorentz factors of $\gamma \sim 5$. Since the ejecta are expected to transition to a slow-acceleration regime once their velocity approaches the fast magnetosonic speed \citep[e.g.,][]{Tchekhovskoy:2009}, for $t \ge t_{\rm rec}$ we remove the magnetization floor in the ejecta region to capture the acceleration process correctly. 

Post reconnection onset, we observe the ejecta to linearly accelerate to $\gamma\sim 14$. This behavior can be understood by assuming conservation of the specific energy flux in the radial direction,
\begin{equation}
\sigma_0 \equiv -\frac{T^{r}_{t}}{\rho u^{r}} \approx \gamma\,(\sigma_{\rm cold}+h+1) \approx \gamma\,\sigma_{\rm cold},
\end{equation}
where $\sigma_{\rm cold} = b^2 / 4\pi\rho c^2$ and the gas enthalpy is $h = (U+p)/\rho$. Here we have a highly magnetized and cold plasma, i.e., $\sigma_{\rm cold} \gg 1$ and $h \ll 1$. The ejecta is launched with a total specific energy flux $\sigma_0 \approx \gamma\sigma_{\rm cold} \approx 2500$, and it linearly accelerates until it reaches the fast-magnetosonic speed, $\gamma = \gamma_{\rm F} = \sigma_{\rm cold,F}^{1/2}$. Assuming conservation of the specific energy flux, we have
\begin{equation}
\sigma_0 \approx \gamma_{\rm F}\,\sigma_{\rm cold,F} \approx \gamma_{\rm F}^{3},
\end{equation}
which gives $\gamma_{\rm F} \approx \sigma_0^{1/3} \approx 14$. The magnetization correspondingly drops to $\sigma_{\rm cold,F} \approx \sigma_0^{2/3} \approx 200$, consistent with the numerical results. Beyond $\gamma \simeq \gamma_{\rm F}$, the ejecta transitions to the slow-acceleration regime, which is analytically predicted to follow $\gamma \propto r^{1/3}$ \citep{Granot:2011}. At larger distances from the magnetar, the magnetic pulse may either collide with the magnetospheric current sheet beyond the light cylinder, or drive a shock in the magnetar wind, both of which are plausible FRB production mechanisms \citep{Beloborodov:2020, Lyubarsky:2020, Lyubarsky:2021, Mahlmann:2022}.

\begin{figure*}
    \centering
    \includegraphics[width=\textwidth,trim= 45mm 0mm 30mm 20mm, clip]{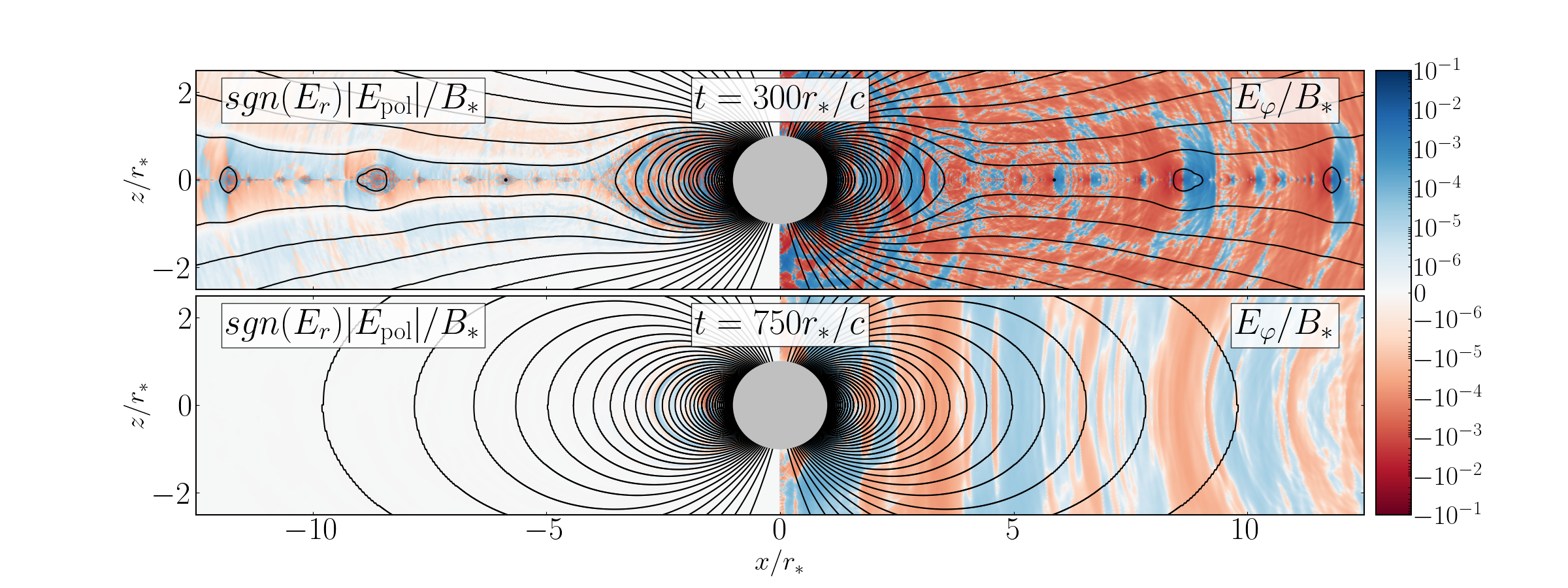}
    \caption{Production of Alfv\'en and fast waves during the eruption. We show electric field components shortly after the onset of reconnection (\textit{top}) and in the post-flare state (\textit{bottom}). The poloidal component, $E_{\rm pol}$, traces Alfv\'en waves, while the toroidal component, $E_{\varphi}$, tracks fast magnetosonic waves. As the field lines in the inner magnetosphere close following the eruption, fast waves gradually escape the magnetosphere, whereas Alfv\'en waves remain confined to the inner region. The electric fields are normalized by the stellar magnetic field strength $B_*$.
    }
    \label{fig:elec}
\end{figure*}

\subsection{\label{sec:AW}Alfv\'en and Fast Wave production in flares}

The inflation of the magnetosphere and the subsequent onset of reconnection generate Alfvén waves (AWs). Their trapping and dissipation in the inner closed magnetosphere may provide additional heat  contributing to the fireball energy and its observed radiation.
We estimate the energy in AWs by approximating it as twice the energy of the poloidal electric field, $\mathcal{E}_{\rm A} \approx 2 \mathcal{E}_{\rm E}^{\rm pol}$. The AW energy stored within the inner magnetosphere turns out to be minuscule, $\mathcal{E}_{\rm A}\ll 1\%\,E_{\rm dip}$. The AWs are continuously erased by grid-scale dissipation as they progressively shear during propagation along closed magnetic field lines.

Collisions between plasmoids generated in the reconnection layer and their bombardment of the closed-field region produce compressive disturbances that launch fast magnetosonic waves (FWs), which show up in the simulation as waves of the toroidal electric field (Fig.~\ref{fig:elec}).
These waves propagate roughly isotropically, and are most prominent during the early reconnection phase, $230 \lesssim t/(r_* c^{-1}) \lesssim 300$. Dissipation of FWs, for example through shock formation \citep{Beloborodov:2023}, could contribute additional energy release. However, similar to AWs, their energy content remains very low, $\ll 1\%\,E_{\rm dip}$. Over time, plasmoid collisions weaken and FWs escape the magnetosphere, causing a decline in FW energy within the inner region. By the end of the simulation, the remaining FW energy is $\mathcal{E}_{\rm F} \sim 2 \mathcal{E}_{\rm E}^{\rm tor} \lesssim 10^{-6} \mathcal{E}_{\rm dip}$ (Fig.~\ref{fig:energetics}b).

\section{\label{sec:summary}Discussion}

In this Letter, we present the first relativistic MHD simulations of an erupting magnetar magnetosphere triggered by crustal shearing of magnetic fields anchored to the stellar surface, motivated by the three giant flares observed to date \citep{Evans:1980, Hurley:1999, Hurley:2005}. Our results show that a substantial fraction of the magnetar’s dipole energy can be released in a magnetically dominated giant plasmoid, whose trailing, hot relativistic plasma could produce the main spike of the gamma-ray flare. The eruption leaves behind hot plasma trapped in the closed magnetosphere. This trapped fireball serves as the energy reservoir powering the extended X-ray emission tail of the flare. Our simulation also demonstrates that the ejecta launches a powerful compressive wave (a fast magnetosonic pulse). This pulse carries sufficient energy to generate FRBs through its interaction with the magnetar wind \citep{Beloborodov:2020} or the magnetospheric current sheet \citep{Lyubarsky:2020,Mahlmann:2022}.

\subsection{Energetics and beaming}
\label{sub:beaming}

\begin{figure}
    \centering
    \includegraphics[width=\columnwidth,trim= 10mm 10mm 10mm 10mm, clip]{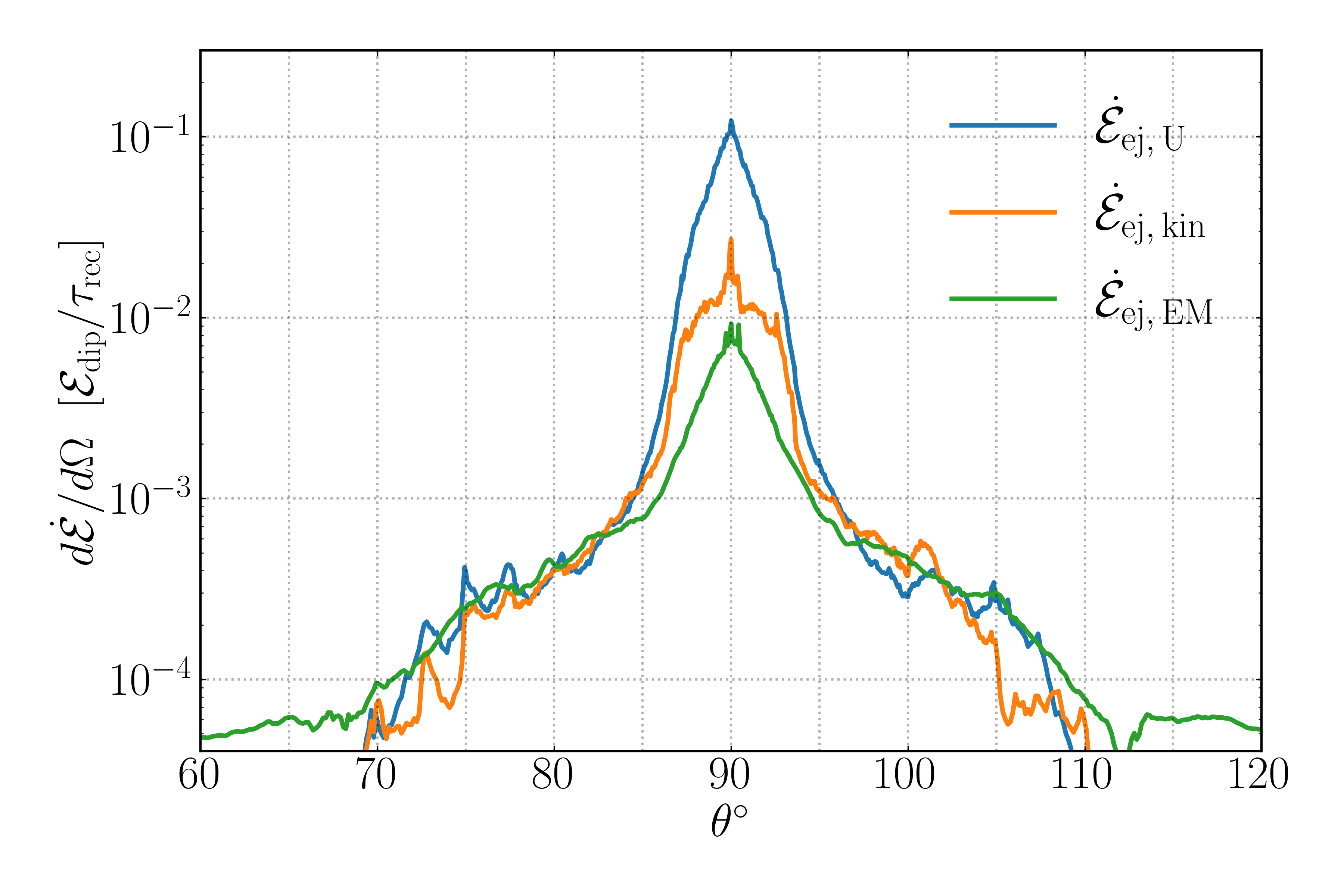}
    \caption{Angular distributions of the thermal, kinetic, and electromagnetic energy fluxes in the ejecta tail per unit solid angle, $d\dot{\mathcal{E}}/d\Omega$, are shown in units of $\mathcal{E}_{\rm dip}/\tau_{\rm rec}$. These quantities are averaged over 20 cells at a radius of $\sim 250\,\rstar$ where the tail is widest at $t = 532r_*/c$ (see Fig.~\ref{fig:bubbletail}). In the ejecta tail, the thermal and kinetic energy fluxes dominate over the electromagnetic flux. In our model, the narrow angular extent of the tail implies non-trivial beaming-factor corrections that must be applied when inferring the intrinsic luminosity of the flare spike from observations.}
    \label{fig:angle}
\end{figure}

For the chosen twist profile, our simulations predict that the total thermal energy carried by the ejecta tail is $\approx 7.7\%\,\mathcal{E}_{\rm dip}$, while the energy stored in the fireball is $\approx 1.5\%\,\mathcal{E}_{\rm dip}$. These efficiencies set the overall energy budgets for the main spike and the extended tail of the flare. The simulations further show that the hot plasma in the ejecta tail moves relativistically ($\gamma \gtrsim 20$) and is predominantly confined to a narrow solid angle, with an angular width of $\sim 20^{\circ}$ (Fig.~\ref{fig:angle}), along the equatorial current sheet. The resulting strong beaming can substantially enhance the apparent luminosity of the initial spike when viewed along the ejection direction, implying that the true released energy may be considerably smaller than its isotropic-equivalent estimate.

Assuming that the giant flare is observed along the ejection direction, the measurement in Fig.~\ref{fig:angle} implies $({d\dot{\mathcal{E}}}/{d\Omega})_{\rm max} \sim 0.1{\mathcal{E}_{\rm dip}}/{\tau_{\rm rec}},
$ and the corresponding isotropic-equivalent thermal energy of the ejecta is

\begin{equation}
\begin{split}
    \mathcal{E}^{\rm max}_{\rm U,\,iso} &\approx 4\pi \left(\frac{d\dot{\mathcal{E}}}{d\Omega}\right)_{\rm max} \times \tau_{\rm rec} \\ &\approx 3.8\times10^{47}\,{\rm erg} \left( \frac{B_*}{10^{15}\,{\rm G}} \right)^2,
\end{split}
\label{eq:Eisomax}
\end{equation}
where we used Eq.~\ref{eqn:edip} for estimating $\mathcal{E}_{\rm dip}$ for a magnetar of radius $\rstar=10\,$km. An offset of the line of sight from the ejection directions leads to substantially lower isotropic-equivalent luminosities, by factors of $10^{-1}$--$10^{-3}$.

Unlike the ejecta, the trapped fireball is quasi-static and its emission is not beamed. Thus, observers with any lines of sight will see a similar total radiation energy emitted by the fireball, $\approx 1.5\%\,\mathcal{E}_{\rm dip} \approx 4.5 \times 10^{45}\,\mathrm{erg}\,(B_*/10^{15}\,\mathrm{G})^{2}$. The fireball energy
will be reduced for flares triggered by
twists located closer to the magnetic poles (see Appendix~\ref{sec:twist_loc}), or for non-axisymmetric twists covering a smaller fraction of the stellar surface. Note that even axisymmetric fireballs will be observed as pulsating if the rotation axis is misaligned with the magnetic axis (since the fireball shape and its radiation are not spherically symmetric). Fully three-dimensional simulations will be required to study more general geometries. 

An intriguing possibility is the observation of
an ``orphan'' pulsating tail emitted by the fireball, without an initial bright spike. This can happen if the line of sight is far from the directions of the thermal ejecta. Future work should refine estimates of the detection probability for the main spike. We expect that magnetar rotation and a more chaotic outflow from the current sheet, arising from three-dimensional effects in reconnection, may lead to a larger opening angle of the thermal ejecta. This would increase the probability of observation of the main spike compared to our axisymmetric model.

\subsection{Comparison with observed giant flares}\label{sec:obs_ener}
The three observed giant flares share a common two-phase structure but differ markedly in energy. Each begins with an extremely bright initial spike lasting $\sim0.2$–$0.5\,$seconds, followed by a softer, pulsating tail of $\sim200$–$400\,$seconds. The 1979 SGR 0526–66 and 1998 SGR 1900+14 events were broadly similar in scale: their initial spikes released $\sim10^{44}$–$10^{45}\,$erg (isotropic equivalent), and their tails carried $\sim10^{44}\,$erg \citep[e.g.,][]{Feroci:2001, Mereghetti:2008}. By contrast, the most recent giant flare detected in 2004 from SGR~1806--20 
was exceptionally energetic: the isotropic-equivalent energy of its initial spike exceeded $\sim 10^{46}\,\mathrm{erg}$, nearly two orders of magnitude larger than in previous events, while its duration was comparable to that of 
the earlier flares. The total energy in the pulsating tails of all three flares was comparable, at $\sim 10^{44}\,\mathrm{erg}$.

The total thermal energy stored in the fireball in our simulations exceeds the energy release inferred from the pulsating tails of observed giant flares. This discrepancy likely has a geometrical origin, as discussed in Sect.~\ref{sub:beaming}; for example, the twisted region may lie closer to the magnetic poles, or it may occupy a smaller fraction of the surface than assumed in our axisymmetric model. 

The large differences in isotropic-equivalent energy release for the main spikes of the three flares can be explained by the beaming effects. This suggests that the 2004 event produced ejecta closer to the line of sight compared to the other two giant flares.\footnote[5]{This is in some tension with the non-detection
of an FRB from the 2004 flare \citep{Tendulkar16}. It suggests either a failed FRB or a missed FRB that was emitted with stronger beaming compared to the gamma-rays.} It is also possible that the 2004 flare was powered by a different physical mechanism altogether. \cite{Bransgrove:2025} have recently proposed a new mechanism of ``crustal eruption'' --- ejection of a magnetic loop from the star itself, as a result of ambipolar diffusion in its interior. This scenario can explain the observed presence of baryonic ejecta \citep{Gaensler05,Granot06}, whereas our magnetospheric eruption model naturally produces an ultra-relativistic baryon-free outflow.

\subsection{Duration of the main spike}
The characteristic dynamical timescale of the flare in our simulations is set by the reconnection time,
\begin{equation}
\tau_{\rm rec} \approx \frac{r_{\rm rec}}{v_{\rm rec}} \approx 10~\mathrm{ms} 
\left(\frac{r_{\rm rec}}{10 r_*}\right)\left(\frac{0.03}{v_{\rm rec}/c}\right),
\label{eq:taurec}
\end{equation}
which is considerably shorter than the observed duration of the main spike, $\sim 250$--$500~\mathrm{ms}$. 
Although the hot ejecta tail requires a few$\,\times\, \tau_{\rm rec}$ to traverse a surface at $\sim 100 r_*$, 
our simulated flare remains significantly shorter. Positioning the twist region closer to the magnetic pole could partially increase the reconnection timescale, $\propto r_{\rm rec} \sim r_C$, where $r_C$ is the equatorial radius of the twisted field line. However, this would also reduce the amount of released energy, which scales approximately as $r_C^{-5/2}$, as discussed in Appendix~\ref{sec:twist_loc}.

Considerable uncertainty remains regarding the physical mechanisms that set the reconnection rate in flare current sheets near the magnetar. Global GR-particle-in-cell simulations of collisionless pair plasmas around black holes find local reconnection rates of $v_{\rm rec}\sim0.1c$ \citep{Bransgrove:2021,El_Mellah:2023,Meringolo:2025}, substantially larger than the rate measured in our ideal MHD simulations. If similarly fast reconnection occurred in magnetar flares, it would shorten the flare duration, $\tau_{\rm rec}$, and increase the flare luminosity. Strong radiative cooling appears to weakly affect the collisionless reconnection rate \citep[e.g.,][]{Sironi:2020}. However, reconnection in the flare current sheet near a magnetar is expected to occur in a deeply radiative regime, where copious pair creation may produce an optically thick, collisional plasma \citep{Uzdensky:2011,Beloborodov:2021}. Reconnection in this regime can be influenced by the strong radiation pressure that develops around the layer, potentially reducing the reconnection rate. The details of this process remain unknown and require first-principles radiative kinetic and MHD simulations.

As the first MHD simulation of a magnetar giant flare, this work adopts three principal simplifications: axisymmetry, an initially dipolar magnetar field, and flat spacetime, to isolate the launching, global dynamics, and energetics of the eruption. Axisymmetry excludes dissipation associated with intrinsically three-dimensional effects, such as kink instabilities of twisted flux tubes, which may alter the outflow geometry and detailed energetics \citep[as seen in FFE simulations;][]{Carrasco:2019,Mahlmann:2023}. Nevertheless, these studies show that global eruptions remain a generic outcome of strongly twisted magnetospheres. Our model is therefore most directly applicable to flares releasing $\lesssim0.2\mathcal{E}_{\rm dip}$ and triggered beyond a few stellar radii, where the field remains approximately dipolar and general-relativistic corrections are negligible.\footnote[6]{For a canonical neutron star, spacetime curvature enhances the polar surface magnetic field by approximately $50\%$ for a fixed asymptotic dipole moment, although this correction decreases to only $\sim10\%$ by $r\sim4\,r_\star$ \citep{Wasserman:1983,Rezzolla:2001_NS_GR}. Conversely, for the fixed local surface-field normalization adopted here, the magnetic field at these radii is approximately $25\%$ weaker than that of the corresponding flat-space dipole. GR can also modify the near-surface field geometry and polar-cap electrodynamics in the rotating magnetosphere \citep{Philippov:2015,Petri:2018}, but is not expected to change the qualitative large-scale eruption mechanism.} 

The dipolar geometry is also important for the present hybrid FFE-MHD treatment, as it produces a single equatorial current sheet whose boundary can be tracked using a prescribed dependence on radius and polar angle (Appendix~\ref{sec:FFE}). Preliminary simulations with higher-multipole fields produce current-sheet and trapped-fireball structures that depend sensitively on the magnetic topology and twisting geometry. While an extended twisted flux bundle may still form a single dominant sheet, more complex configurations can generate moving or off-equatorial dissipative regions that the present geometry-dependent prescription cannot track reliably, requiring an FFE-MHD switching criterion based only on local plasma properties.
Together, these assumptions provide a controlled description of the large-scale eruption regime studied here, while more energetic events originating deeper in the magnetosphere will require fully 3D GRMHD simulations with multipolar geometries.

\begin{acknowledgments}

We thank the anonymous referee for their constructive suggestions that greatly improved the text. KC thanks Jens Mahlmann and Christopher Thompson for many interesting discussions. This work was supported by NASA grant 80NSSC22K1054 (KC and AP), Simons Foundation (00001470, AP and BR), and facilitated by Multimessenger Plasma Physics Center (MPPC, AP and AMB), NSF grant No. PHY-2206607. KC is also supported by a CITA Postdoctoral Fellowship (NSERC DIS fund 513671). AP additionally acknowledges support by an Alfred P. Sloan Fellowship, and a Packard Foundation Fellowship in Science and Engineering. AMB and ERM acknowledge support from NASA's ATP program under grant 80NSSC24K1229.
AMB is also supported by NSF grant AST-2408199 and Simons Foundation award 446228. 
ERM acknowledges support by the National Science Foundation under grants No. PHY-2309210 and AST-2307394.  
BR is supported by the Natural Sciences \& Engineering Research Council of Canada (NSERC) [funding reference number 568580], and the Canadian Space Agency (23JWGO2A01). This research was supported in part by grant NSF PHY-2309135 to the Kavli Institute for Theoretical Physics (KITP). KP acknowledges support from the Laboratory Directed Research and Development Program at Princeton Plasma Physics Laboratory, a national laboratory operated by Princeton University for the U.S.\ Department of Energy under Prime Contract No.\ DE-AC02-09CH11466.

This research is part of the Frontera computing project \citep{Frontera} at the Texas Advanced Computing Center (LRAC-AST21006). Frontera is made possible by NSF award OAC-1818253. Part of the simulations were performed on DOE OLCF Summit under
allocations AST198. This research used resources of the Oak Ridge Leadership Computing Facility at the Oak Ridge National Laboratory, which is supported by the Office of Science of the U.S. Department of Energy under Contract No. DE-AC05-00OR22725.

\end{acknowledgments}

\appendix
\vspace{-3mm}
\section{Methodology for hybrid FFE-MHD evolution}
\label{sec:FFE}

We evolve the ideal MHD equations throughout the domain and impose force-free (FF) conditions where appropriate by enforcing thresholds on the magnetization and plasma-$\beta$, and by suppressing the fluid velocity along the magnetic field ($v_{\parallel}=0$ close to the star), following \citet{Parfrey:2017,Parfrey:2024}. Coupling between the FF and MHD regions is achieved through a controlled buffer zone evolved with the entropy equation, as discussed below. This hybrid FF–MHD treatment yields behavior consistent with approaches that explicitly evolve the FF equations rather than enforcing floors \citep{Chael:2024,Komissarov:2025}. Furthermore our high grid resolution ensures that we obtain converged plasmoid-mediated reconnection rates similar to resolved resistive MHD simulations due to the small numerical resistivity in the asymptotic regime in our ideal MHD simulations \citep{Grehan:2025}.

To stably evolve the magnetospheric twist, reconnection in the current sheet, and the acceleration of the ejecta, we adopt a region-dependent primitive-variable recovery strategy, switching between the energy and entropy evolution equations in our GRMHD code \hammer{} \citep{Liska:2022:hamr}. Standard GRMHD schemes conserve total energy and impose density and pressure floors in highly magnetized regions, such as relativistic jets. In the initial phase of the twist buildup, the magnetosphere is strongly magnetically dominated. We therefore employ energy inversion throughout the domain, applying a density floor that maintains a fixed cold magnetization of $\sigma_{\rm cold}=500$ and an internal-energy floor that enforces $\beta \approx 10^{-8}$, effectively approximating force-free conditions. This approach ensures that the large reservoir of electromagnetic energy generated by twist injection is evolved self-consistently. The energy-inversion method becomes unreliable when the internal energy is extremely small, because it is obtained as the difference between the conserved total energy and the electromagnetic energy. When these two large quantities nearly cancel, truncation errors can yield unphysical values, including negative pressures. To avoid this, we tie the density and internal-energy floors explicitly to the local magnetic-field strength. In addition, near the star, we suppress the velocity component parallel to the magnetic field, $v_{\parallel}$, to prevent gas accumulation along field lines as the magnetosphere inflates, $t \gtrsim t_{\rm inf}$.

\begin{figure*}
    \centering
    \includegraphics[width=\textwidth,trim= 50mm 0mm 50mm 20mm, clip]{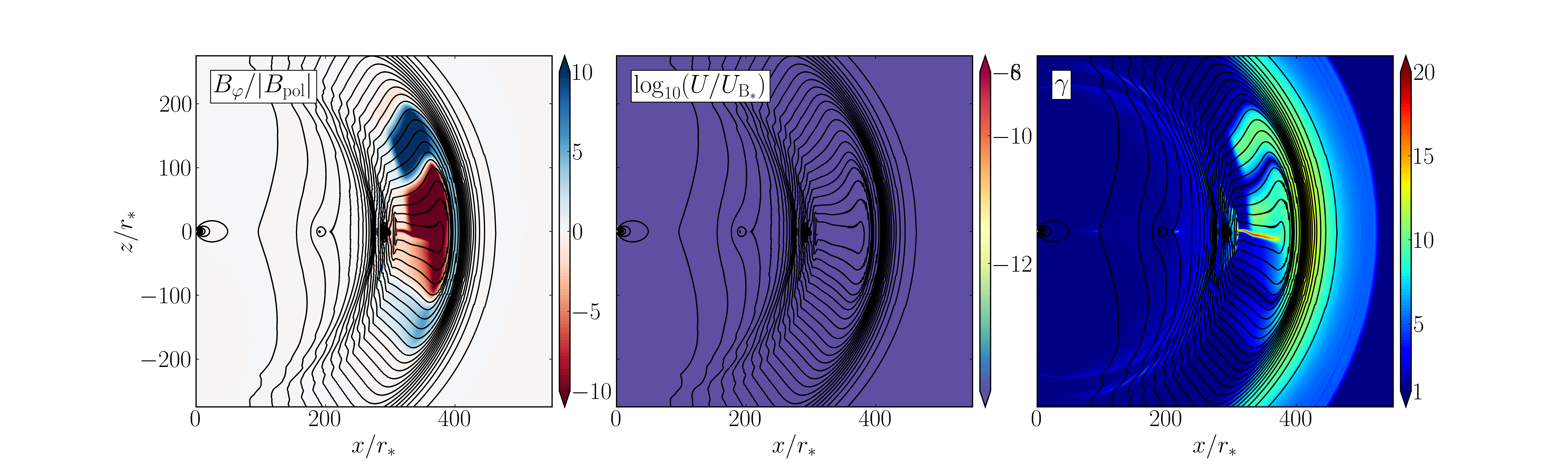}
    \caption{The same as Fig.~\ref{fig:bubbletail}, but evolved using a force-free prescription. We find that our MHD simulation accurately captures the dynamics and magnetic-field distribution of the ejecta head and the magnetic pulse when compared to the floored FFE version of the simulation. The FFE-only simulation does not capture reconnection-heating as expected. 
    }
    \label{fig:FFE_bubble}
\end{figure*}

A different strategy is required once reconnection begins and during the subsequent post-reconnection evolution, $t \gtrsim t_{\rm rec}$. To determine where and when reconnection first occurs, we perform a simulation at the same resolution as the fiducial run but enforcing our force-free condition throughout the domain and over the entire eruption evolution. In this setup, during reconnection, thermal energy is continually reset to its floor value, causing the current sheet to lose pressure support and the domain to lose energy over time. Nevertheless, the model accurately captures the reconnection onset time, the current-sheet geometry, and the large-scale ejecta morphology. As shown in Fig.~\ref{fig:FFE_bubble}, the overall ejecta structure closely matches that of the fiducial run, apart from the absence of hot plasmoids. Importantly, because the ejecta head and the forward-compressed pulse remains cold and accelerates, the floored FFE simulation provides a robust estimate of the plasma acceleration before it reaches the fast-magnetosonic speed.

Guided by our force-free results, we activate a hybrid primitive-variable recovery strategy at $t=t_{\rm rec}$, switching between energy- and entropy-inversion depending on the plasma conditions. We partition the simulation domain into four regions: (i) the equatorial reconnection layer and trailing ejecta tail, which contain hot relativistic plasma; (ii) the ejecta head, consisting of cold, accelerating gas; (iii) the fireball, containing quasi-steady hot plasma; and (iv) the inflated magnetosphere, containing sub-relativistic cold gas threaded by non-twisted magnetic fields.

The hot equatorial ejecta, where magnetic dissipation produces relativistic temperatures and strong pressure gradients, are evolved using the energy equation without applying floors, enabling accurate tracking of thermal-energy generation and bulk acceleration. The ejecta head is also evolved with the energy equation but retains only internal-energy floors, allowing its magnetization to decrease steadily as it accelerates. Removing the magnetization floor is essential: the acceleration naturally slows as the flow approaches the fast-magnetosonic limit, analogous to the behavior of relativistic jets (Sec.~\ref{sec:bubble}). In contrast, the FFE-only model artificially maintains high magnetization and therefore produces continuous linear acceleration beyond the expected fast-speed limit.

The fireball and the inflated magnetosphere are evolved using the entropy equation. Alongside total-energy conservation, \hammer{} evolves the entropy equation $\nabla_{\mu}(\kappa \rho u^{\mu}) = 0,$ where the specific entropy is given by $s = (\Gamma_{\rm ad}-1)\log{\kappa}$.
This entropy variable is used as a backup primitive-recovery method \citep{Noble:2009} when the standard inversion fails, typically in highly magnetized regions \citep{Chatterjee:2019}. The entropy-inversion method yields a lower estimate of the internal energy, which is subdominant to the magnetic energy in these regions, discounting any heating due to reconnection or shocks. Previous studies have shown that entropy switching performs well for problems involving magnetized flows, including spherical accretion \citep{Porth:2017}. However, the entropy scheme cannot reliably capture acceleration or reconnection heating, making it unsuitable for the current sheet and the ejecta. It does, nonetheless, accurately describe advection of magnetic fields and the transport of both hot and cold plasma, making it well suited for evolving the fireball and the cold magnetosphere while providing a controlled buffer region between the reconnecting layer and the surrounding cold plasma. The density and internal-energy floors used during the initial twist evolution are removed in these regions, but we continue to suppress $v_{\parallel}$ in cold plasma with non-twisted magnetic fields to avoid density pile-up and to maintain the floor magnetization and plasma-$\beta$.

The choice between switching between energy and entropy evolution is made using a smooth, local switching criterion. We adopt entropy inversion when
\begin{equation}
    f_{\rm ent}> 0.5\times \left[ 1-\left(\frac{\rstar}{r}\right)^8\right] + 0.48\times (\sin{\theta})^8,
\end{equation}
where $f_{\rm ent}:=(b^2/8\pi)/(\rho c^2+U+p+b^2/8\pi)$ quantifies magnetic dominance. The radial dependence identifies the fireball region and preserves a cold, non-twisted magnetosphere interior to it, while the strong $(\sin\theta)^8$ dependence cleanly isolates the equatorial current sheet while still allowing extended hot plasmoids to be handled by energy inversion. The ejecta region is tracked with a moving surface at $r \approx 0.98\, c t$, with a correction for the location of the ejecta head at the time of onset of reconnection. We also introduce entropy inversion in a small buffer zone between the hot ejecta tail and the ejecta head to permit limited mixing without disrupting the acceleration physics.

We validated this hybrid inversion/floor scheme by confirming that the total energy inside $r \le 30\,\rstar$ is conserved with an error $\lesssim 0.1\%$ of the initial dipole energy. At larger radii, discrepancies of a few percent arise due to the difficulty of capturing dissipation accurately in the transition zone connecting the current sheet, ejecta tail, and ejecta head, as well as from dissipation in the magnetic pulse associated with limited grid resolution.

\section{\label{sec:twist_loc}Dependence on the twist location}

The burst luminosity and duration, $\sim \tau_{\rm rec}$, depend on the location of the imposed twist on the stellar surface, which is not known. To quantify the dependence of the burst luminosity on the twist location, we performed three $10080^2$ floored FFE simulations with different choices of the twist colatitude, $\theta_{\rm C} = 25^\circ$, $35^\circ$, and $45^\circ$, respectively. During the initial twisting phase, $\psi \lesssim 1\,{\rm rad}$, the injected energy is predominantly converted into the growth of the toroidal magnetic field. Following \cite{Beloborodov:2009} and \cite{Parfrey:2013}, the excess energy 
stored in the twisted magnetosphere can be estimated, in the small-twist limit 
$\psi \ll 1$, as:
\begin{equation}
    \mathcal{E}_{\rm twist}(t)\approx \int\frac{B_{\varphi}^2}{8\pi}dV=\frac{\mu}{2c\rstar}\int^{\pi}_0 I\psi\sin(2\theta)\,d\theta,
    \label{eq:Etwist_approx}
\end{equation}
where we express the magnetic moment $\mu$ in terms of the initial dipole energy,
$\mathcal{E}_{\rm dip} = \mu^{2}/(3\rstar^{3})$. We also make use of the poloidal 
current, $I = (1/2)c B_{\varphi} r \sin\theta$, obtained via Stokes' 
theorem. Expressing the twist angle $\psi$ as the azimuthal displacement of a field line, integrated along its length (see Appendix A of \citealt{Beloborodov:2009}), we can relate $I$ to $\psi$ as:
\begin{equation}
    I=\frac{c\,\mu\, \psi(t,\theta)}{4\rstar^2}\frac{\sin^4\theta}{\cos\theta}.
\end{equation}
Replacing $I$ and $\mu$ in Eqn.~\ref{eq:Etwist_approx},
\begin{equation}
    \mathcal{E}_{\rm twist}(t)\approx\frac{3}{4}\mathcal{E}_{\rm dip}\int_0^{\pi} \psi^2(t,\theta)\sin^5\theta\, d\theta,
\end{equation}
where the total twist angle $\psi(t, \theta)\approx \omega(\theta)t$ (Eqn.~\ref{eqn:twist}) in the initial spin-up phase. Assuming a thin shearing ring, $\theta_{\rm T}\ll1$, we get:
\begin{equation}
    \mathcal{E}_{\rm twist}^{\rm shear}(t)\approx \frac{3}{4}\mathcal{E}_{\rm dip}\psi^2(t)\theta_{\rm T}\sin^5\theta_{\rm C}.
    \label{eq:Etwist_thin}
\end{equation}

Using our three FFE-only simulations, we estimate the twist energies at 
$\psi \sim 0.5$ from Eq.~\ref{eq:Etwist_thin} as 
$\mathcal{E}_{\rm twist}^{\rm shear}(\theta_{\rm C}) \approx 3\%\,\mathcal{E}_{\rm dip}\,\sin^{5}\theta_{\rm C}$, 
keeping $\theta_{\rm T}=0.05\pi$ fixed across models. For 
$\theta_{\rm C} = (25^{\circ},\,35^{\circ},\,45^{\circ})$, this yields 
$\mathcal{E}_{\rm twist}^{\rm shear} \approx (0.04\%,\,0.18\%,\,0.53\%) \times \mathcal{E}_{\rm dip}$, which we 
compare with the corresponding values measured directly from the simulations: 
$(0.06\%,\,0.18\%,\,0.43\%) \times \mathcal{E}_{\rm dip}$. We find that this expression 
provides a good approximation to the twist energy up to 
$\psi \sim 0.5\,$rad. As expected from our analytic estimate, the twist energy 
decreases as the twist center moves closer to the pole (smaller $\theta_{\rm C}$). The magnetic flux 
function for a dipole field is 
$\Psi = (2 \pi \mu / r) \sin^2 \theta$, and so, $\sin^2\theta / r$ remains 
constant along a field line. Consequently, the twist energy scales as 
$\mathcal{E}_{\rm twist}^{\rm shear} \propto \theta_{\rm T} r_{\rm C}^{-5/2}$, where 
$r_{\rm C} = \rstar / \sin^2\theta_{\rm C}$ is the equatorial radius of the 
field line whose footpoint is at $\theta_{\rm C}$. 

For $\psi \gtrsim 1$, the magnetosphere enters the inflationary stage, with a 
portion of the injected energy contributing to the growth of the poloidal field (Fig.~\ref{fig:energetics}b). Empirically, we find that the total twist energy injected up to reconnection, 
$\mathcal{E}_{\rm twist}(t_{\rm rec})$, is approximately $40\,\mathcal{E}_{\rm twist}(\psi \sim 0.5)$ 
for our chosen shearing prescription. We also observe that the exact reconnection time (and thus $\mathcal{E}_{\rm twist}$ at $t_{\rm rec}$) increases slightly with grid resolution, since the equatorial magnetic field can become increasingly monopolar before undergoing reconnection at the grid scale (which can only be avoided by introducing an explicit resistive length scale). Twists located closer to the pole produce larger twisted loops (i.e., larger 
$r_{\rm C}$), trigger reconnection at greater radii, and release a smaller 
amount of energy, scaling approximately as $\sim \mathcal{E}_{\rm twist} \sim r_{\rm C}^{-5/2}$.

\begin{figure}
    \centering
    \includegraphics[width=0.7\textwidth,trim= 20mm 0mm 20mm 20mm, clip]{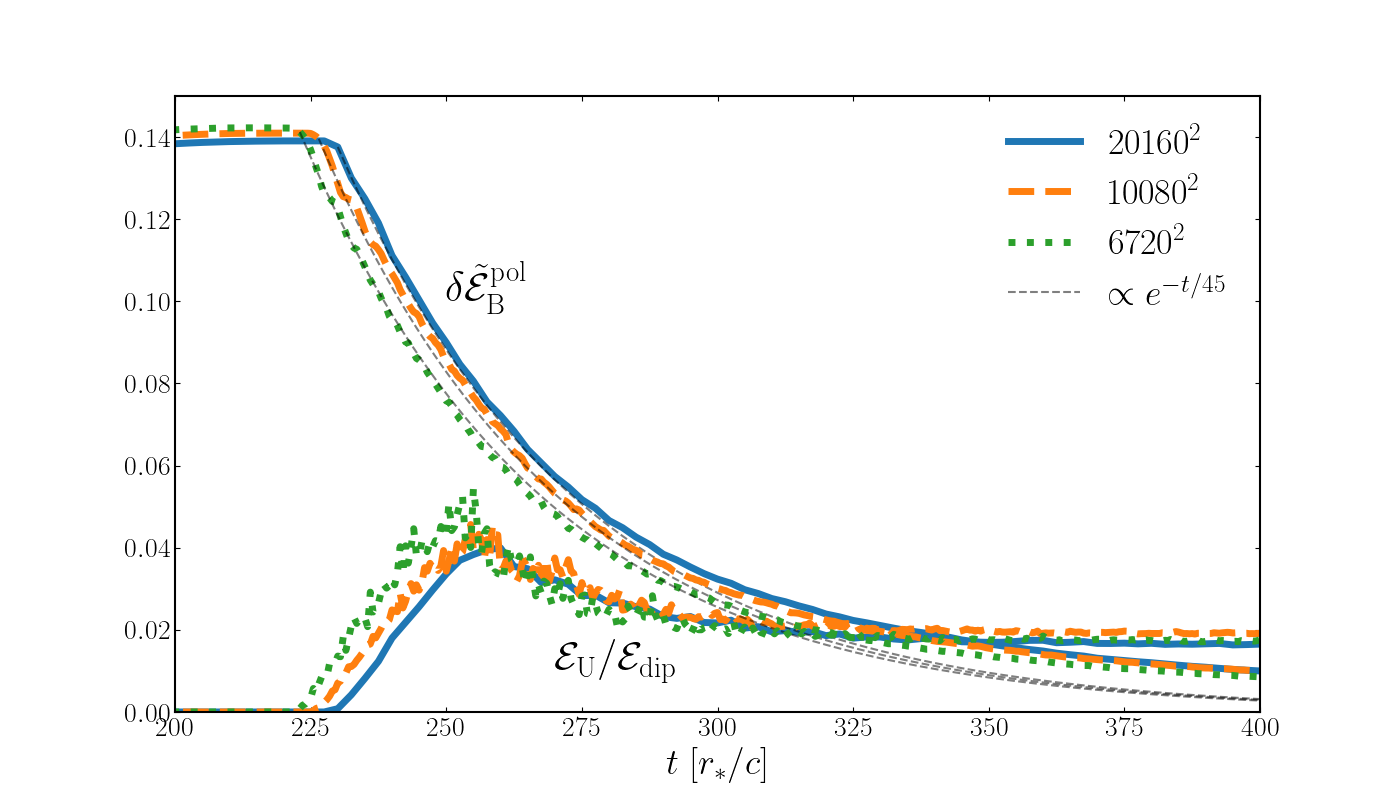}
    \caption{Resolution dependence of the energetics in the twisted magnetosphere for three effective grid resolutions, $6720^2$, $10080^2$, and $20160^2$. The timescale of the reconnection-driven decay of the poloidal magnetic energy stored in the inner magnetosphere, $\mathcal{E}_{\rm B}^{\rm pol}$ (Fig.~\ref{fig:energetics}b), is $\tau_{\rm rec}\approx 90\rstar/c$ in all three simulations, indicating a converged reconnection rate (see Sec.~\ref{sec:fireball}). The thermal energy of the trapped fireball, $\mathcal{E}_{\rm U}$ (Fig.~\ref{fig:energetics}a), also shows consistent behavior at these resolutions.
    }
    
    \label{fig:converge}
\end{figure}

\section{\label{sec:convergence}Numerical resolution dependence of the energetics}

We perform two additional hybrid FFE--MHD simulations at effective grid resolutions of $6720^2$ and $10080^2$ and compare them with the fiducial $20160^2$ simulation in Fig.~\ref{fig:converge}. These calculations are intended to assess the sensitivity of the global eruption dynamics and energetics to grid resolution. The simulations evolve the ideal MHD equations, reconnection is mediated by numerical dissipation at the grid-scale, instead of an explicit resistive length-scale.

The primary resolution-dependent quantity is the onset time of the initial reconnection event. Reconnection begins when the thickness of the equatorial current sheet becomes comparable to the local grid scale. The higher-resolution simulations therefore allow the magnetosphere to inflate slightly further and exhibit a delayed reconnection onset. This shows that the onset is numerically triggered by the grid scale rather than set by a resolved physical dissipation scale.

However, once reconnection begins, the macroscopic evolution is similar among the three resolutions. The excess poloidal magnetic energy decays approximately as $\delta E_{\rm B}^{\rm pol}\propto \exp(-t/45\,r_\star c^{-1})$,
corresponding to a characteristic reconnection timescale of
$\tau_{\rm rec}\simeq90\,r_\star/c$. The characteristic radius of the initial reconnection events is also similar, with
$\langle r_{\rm rec}\rangle\simeq3.5\,r_\star$. The resulting reconnection rate estimate $v_{\rm rec}\simeq0.038c$ is consistent with previous studies showing that, although the effective resistivity in ideal-MHD simulations is numerical, the reconnection rate reaches an asymptotic value once the current sheet enters the plasmoid-mediated regime \citep{Ripperda:2022, Grehan:2025}.

Figure~\ref{fig:converge} shows that, despite the resolution-dependent shift in reconnection onset, the subsequent magnetic-energy decay and late-time magnetic energy agree to within a few percent across all three simulations. The trapped-fireball energy, $\mathcal{E}_{\rm U}$, shows similarly close agreement, with the slightly larger value in the intermediate-resolution run likely reflecting small differences in the chaotic plasmoid evolution within the current sheet. These results demonstrate numerical consistency of the global post-onset evolution and energetics, rather than formal convergence; establishing the latter will require explicitly resistive relativistic-MHD simulations with a resolved physical dissipation scale. Nevertheless, the close agreement across resolutions shows that the principal eruption dynamics and energetic estimates are robust over the tested range.

\bibliography{magnetar}{}
\bibliographystyle{aasjournalv7}

\end{document}